\documentclass{pasj01}
\Received{$\langle$08-Nov-2015$\rangle$}
\Accepted{$\langle$14-Dec-2015$\rangle$}
\Published{$\langle$-$\rangle$}

\begin{document}

\title{NEAR-INFRARED IMAGING POLARIMETRY OF LkCa 15: A POSSIBLE WARPED INNER DISK\thanks[\dag]{Based on IRCS and HiCIAO data collected at Subaru Telescope, which is operated by the National Astronomical Observatory of Japan.}}

\author{Daehyeon OH\altaffilmark{1,2,*}}
\author{Jun HASHIMOTO\altaffilmark{3}}
\author{Motohide TAMURA\altaffilmark{2,3,4}}
\author{John WISNIEWSKI\altaffilmark{5}}

\author{Eiji AKIYAMA\altaffilmark{2}}
\author{Thayne CURRIE\altaffilmark{6}}
\author{Satoshi MAYAMA\altaffilmark{7}}
\author{Michihiro TAKAMI\altaffilmark{8}}
\author{Christian THALMANN\altaffilmark{9}}

\author{Tomoyuki KUDO\altaffilmark{6}}
\author{Nobuhiko KUSAKABE\altaffilmark{3}}
\author{Lyu ABE\altaffilmark{10}}
\author{Wolfgang BRANDNER\altaffilmark{11}}
\author{Timothy D. BRANDT\altaffilmark{12}}
\author{Joseph C. CARSON\altaffilmark{13}}
\author{Sebastian EGNER\altaffilmark{6}}
\author{Markus FELDT\altaffilmark{12}}
\author{Miwa GOTO\altaffilmark{14}}
\author{Carol A. GRADY\altaffilmark{15,16,17}}
\author{Olivier GUYON\altaffilmark{6}}
\author{Yutaka HAYANO\altaffilmark{6}}
\author{Masahiko HAYASHI\altaffilmark{2}}
\author{Saeko S. HAYASHI\altaffilmark{6}}
\author{Thomas HENNING\altaffilmark{11}}
\author{Klaus W. HODAPP\altaffilmark{18}}
\author{Miki ISHII\altaffilmark{2}}
\author{Masanori IYE\altaffilmark{2}}
\author{Markus JANSON\altaffilmark{19}}
\author{Ryo KANDORI\altaffilmark{2}}
\author{Gillian R. KNAPP\altaffilmark{19}}
\author{Masayuki KUZUHARA\altaffilmark{20}}
\author{Jungmi KWON\altaffilmark{4}}
\author{Taro MATSUO\altaffilmark{21}}
\author{Michael W. MCELWAIN\altaffilmark{15}}
\author{Shoken MIYAMA\altaffilmark{22}}
\author{Jun-Ichi MORINO\altaffilmark{2}}
\author{Amaya MORO-MARTIN\altaffilmark{23,24}}
\author{Tetsuo NISHIMURA\altaffilmark{6}}
\author{Tae-Soo PYO\altaffilmark{6}}
\author{Eugene SERABYN\altaffilmark{26}}
\author{Takuya SUENAGA\altaffilmark{1,2}}
\author{Hiroshi SUTO\altaffilmark{2,3}}
\author{Ryuji SUZUKI\altaffilmark{2}}
\author{Yasuhiro H. TAKAHASHI\altaffilmark{2,4}}
\author{Naruhisa TAKATO\altaffilmark{6}}
\author{Hiroshi TERADA\altaffilmark{2}}
\author{Edwin L. TURNER\altaffilmark{19,26}}
\author{Makoto WATANABE\altaffilmark{27}}
\author{Toru YAMADA\altaffilmark{28}}
\author{Hideki TAKAMI\altaffilmark{2}}
\author{Tomonori USUDA\altaffilmark{2}}

\altaffiltext{1}{Department of Astronomical Science, The Graduate University for Advanced Studies (SOKENDAI), 2-21-1 Osawa, Mitaka, Tokyo, 181-8588, Japan}
\altaffiltext{2}{National Astronomical Observatory of Japan, 2-21-1, Osawa, Mitaka, Tokyo, 181-8588, Japan}
\altaffiltext{3}{Astrobiology Center of NINS, 2-21-1, Osawa, Mitaka, Tokyo, 181-8588, Japan}
\altaffiltext{4}{Department of Astronomy, The University of Tokyo, 7-3-1, Hongo, Bunkyo-ku, Tokyo, 113-0033, Japan}
\altaffiltext{5}{H. L. Dodge Department of Physics \& Astronomy, University of Oklahoma, 440 W Brooks St Norman, OK 73019, USA}
\altaffiltext{6}{Subaru Telescope, National Astronomical Observatory of Japan, 650 North A'ohoku Place, Hilo, HI96720, USA}
\altaffiltext{7}{The Center for the Promotion of Integrated Sciences, The Graduate University for Advanced Studies (SOKENDAI), Shonan International Village, Hayama-cho, Miura-gun, Kanagawa 240-0193, Japan}
\altaffiltext{8}{Institute of Astronomy and Astrophysics, Academia Sinica, P.O. Box 23-141, Taipei 10617, Taiwan}
\altaffiltext{9}{Swiss Federal Institute of Technology (ETH Zurich), Institute for Astronomy,Wolfgang-Pauli-Strasse 27, CH-8093 Zurich, Switzerland}
\altaffiltext{10}{Laboratoire Lagrange (UMR 7293), Universite de Nice-Sophia Antipolis, CNRS, Observatoire de la Coted'azur, 28 avenue Valrose, 06108 Nice Cedex 2, France}
\altaffiltext{11}{Max Planck Institute for Astronomy, K\"{o}onigstuhl 17, 69117 Heidelberg, Germany}
\altaffiltext{12}{Astrophysics Department, Institute for Advanced Study, Princeton, NJ, USA}
\altaffiltext{13}{Department of Physics and Astronomy, College of Charleston, 58 Coming St., Charleston, SC 29424, USA}
\altaffiltext{14}{Universitats-Sternwarte Munchen, Ludwig-Maximilians-Universitat, Scheinerstr. 1, D-81679 Munchen,Germany}
\altaffiltext{15}{Exoplanets and Stellar Astrophysics Laboratory, Code 667, Goddard Space Flight Center, Greenbelt, MD 20771, USA}
\altaffiltext{16}{Eureka Scientific, 2452 Delmer, Suite 100, Oakland CA96002, USA}
\altaffiltext{17}{Goddard Center for Astrobiology}
\altaffiltext{18}{Institute for Astronomy, University of Hawaii, 640 N. A'ohoku Place, Hilo, HI 96720, USA}
\altaffiltext{19}{Department of Astrophysical Science, Princeton University, Peyton Hall, Ivy Lane, Princeton, NJ08544, USA}
\altaffiltext{20}{Department of Earth and Planetary Sciences, Tokyo Institute of Technology, 2-12-1 Ookayama, Meguro-ku, Tokyo 152-8551, Japan}
\altaffiltext{21}{Department of Astronomy, Kyoto University, Kitashirakawa-Oiwake-cho, Sakyo-ku, Kyoto, Kyoto 606-8502, Japan}
\altaffiltext{22}{Hiroshima University, 1-3-2, Kagamiyama, Higashihiroshima, Hiroshima 739-8511, Japan}
\altaffiltext{23}{Space Telescope Science Institute, 3700 San Martin Drive, Baltimore, MD 21218, USA}
\altaffiltext{24}{Center for Astrophysical Sciences, Johns Hopkins University, Baltimore MD 21218, USA}
\altaffiltext{25}{Jet Propulsion Laboratory, California Institute of Technology, Pasadena, CA, 171-113, USA}
\altaffiltext{26}{Kavli Institute for Physics and Mathematics of the Universe, The University of Tokyo, 5-1-5, Kashiwanoha, Kashiwa, Chiba 277-8568, Japan}
\altaffiltext{27}{Department of Cosmosciences, Hokkaido University, Kita-ku, Sapporo, Hokkaido 060-0810, Japan}
\altaffiltext{28}{Astronomical Institute, Tohoku University, Aoba-ku, Sendai, Miyagi 980-8578, Japan}

\email{daehyun.oh@nao.ac.jp}
\maketitle

\begin{abstract}
  We present high-contrast H-band polarized intensity images of the transitional disk around the young solar-like star LkCa 15. By utilizing Subaru/HiCIAO for polarimetric differential imaging, both the angular resolution and the inner working angle reach 0.07\arcsec and r=0.1\arcsec, respectively. We obtained a clearly resolved gap (width $\lesssim$ 27 AU) at $\sim$ 48 AU from the central star. This gap is consistent with images reported in previous studies. We also confirmed the existence of a bright inner disk with a misaligned position angle of 13$\pm$4$\degree$ with respect to that of the outer disk, i.e., the inner disk is possibly warped. The large gap and the warped inner disk both point to the existence of a multiple planetary system with a mass of $\lesssim$1$M_{\rm Jup}$.

\end{abstract}

\section{Introduction}

  The circumstellar disks around young stars are the main birthplace of giant gas planets. Analysis of their spectral energy distribution (SED) and the results of interferometry at infrared to millimeter wavelengths reveal the evidence of gap and cavity structures in many circumstellar disks. Such disks have been called transitional disks, and are thought to be an intermediate phase between gas-rich primordial disks and gas-poor debris disks (e.g., \cite{esp14}). When newly formed planet(s) are embedded in the disks, a gap structure (i.e., optically thick inner and outer disks separated by an optically thin gap) instead of a cavity (i.e., a complete lack of an inner disk) is predicted to form by disk-planet interactions \citep{kle12}. Therefore, disks with a gap structure could indicate the birth of the giant gas planets. Hence, understanding the detailed structures of the transitional disks could unveil the origin of our planetary system.
  
  The progress of high-contrast imaging in the last decade allows us to see more details in the transitional disks. By direct imaging, the incredible diversity of the disk morphology, such as spiral and gap structures, has been revealed (e.g., \cite{has12}). LkCa 15 (K5, 0.97 {\it M}$_{\odot}$, 2$-$5 Myr old; \cite{sim00}), a young solar-like star located in the Taurus-Auriga region ($\sim$140 pc), is one of the most intensively studied transitional disk systems around the T Tauri star. \citet{esp07} conducted a detailed analysis of SED in LkCa 15 and suggested the existence of a gap structure at $\sim$46AU. 
Such a gap structure has been confirmed by millimeter (mm) interferometry (\cite{pie06}) and near-infrared (NIR) high-contrast direct imaging (\cite{tha10}).

  Subsequently, protoplanet candidates LkCa 15 b and c ($\lesssim$5$\sim$10$M_{\rm Jup}$) were discovered (\cite{kra12,sal15}). More recently, the inner disk was newly discovered by \citet{tha15} at optical wavelengths (590-890 nm). Therefore, LkCa 15 may serve an excellent laboratory for studying the interaction between infant planets and the protoplanetary disk structure they sculp.

  Here, we present the results of new high-contrast NIR (1.6$ \mu$m) polarization imaging carried out on the LkCa15 disk. The combination of the High Contrast Instrument for the Subaru Next Generation Adaptive Optics(HiCIAO; \cite{tam06}) and Polarimetric Differential Imaging (PDI) provides a high-contrast image that is unprecedented in quality at infrared wavelengths and enables us to both clearly confirm and quantitatively analyze the wide gap structure and the inner disk. We report the warped inner disk and discuss the potential origin of a gapped and warped disk around LkCa 15. The gapped and warped disk suggests the existence of a multiple planetary system.

\section{OBSERVATIONS AND DATA REDUCTION}

\begin{figure}
\begin{center}
\FigureFile(80mm,){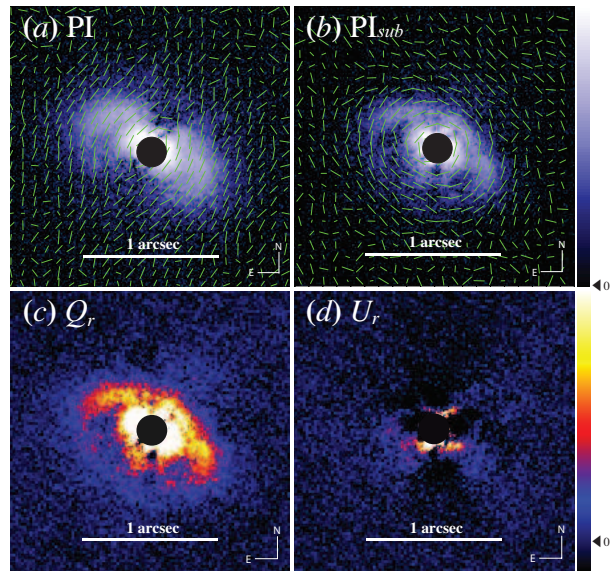}
\end{center}
\caption{ {\it Top}:  PI and overlapped polarization vector map images ($2.0\arcsec\times2.0\arcsec$) before ({\it a}) and after ({\it b}) halo subtraction. The saturated region is occulted by a software mask (r$\sim$0.1\arcsec), the vectors are binned with spatial resolution, and the lengths are arbitrary for presentation purposes. ({\it a}): The effect of a polarized halo appears to have a tendency toward the minor axis of the disk. ({\it b}): The polarization tendency to the minor axis was removed, and the disk-origin polarization along the disk surface was revealed. {\it Bottom}: The radial Stokes {\it Q}$_{\it r}$ ({\it c}) and {\it U}$_{\it r}$ ({\it d}) images. In the {\it Q}$_{\it r}$ image, both the outer and inner disks are significantly detected as expected from the PI image. On the other hand, the {\it U}$_{\it r}$ image shows no disk-like component.}\label{piqrur}
\end{figure}

  The PDI observations of LkCa 15 were performed in the {\it H}-band on 2013 Nov 22 with HiCIAO on the Subaru Telescope combined with AO188 \citep{hay10}. Each image has a 5\arcsec$\times5\arcsec$ field of view (FOV) with a pixel scale of 9.5 mas/pixel. We obtained 17 data sets with 30 s exposure. The total integration time on the source of the polarization intensity image was 2040 s. All of our observations were conducted under the program of the SEEDS (Strategic Explorations of Exoplanets and Disks with Subaru; \cite{tam09}) project.

  The polarimetric data were reduced in the standard manner of infrared image reduction that uses the custom IRAF\footnote{The IRAF software is distributed by the National Optical Astronomy Observatory, which is operated by the Association of Universities for Research in Astronomy (AURA) under a cooperative agreement with the National Science Foundation.} pipeline designed by \citet{has11}. The Stokes {\it Q} and {\it U} images were obtained by the standard method for differential polarimetry \citep{hin09}. The polarized intensity (PI) image was obtained as $({\rm\it Q}^2+{\rm\it U}^2)^{1/2}$. Because the convolved point spread function (PSF) cannot be perfectly removed by standard procedures, a residual stellar halo was sometimes observed in the obtained PI images. To remove the effect of this polarized halo, we constructed the polarization halo model by using the derived average polarization strength (0.67$\pm$0.03$\%$) and average polarization angle (149.1$\pm$0.5$\degree$), and subtracted this from the Stokes {\it Q} and {\it U} images. From the halo-subtracted Stokes {\it Q}$_{sub}$ and {\it U}$_{sub}$ images, the final halo-subtracted PI$_{sub}$ image were generated (Figure \ref{piqrur}b). To verify this result, we converted the coordinate system of Stokes {\it Q} and {\it U} to the radial Stokes {\it Q}$_{\it r}$ and {\it U}$_{\it r}$ (\cite{ave14}), because the Stokes {\it Q}$_{\it r}$ image must show scattering polarization similar to the PI$_{sub}$ image, while the Stokes {\it U}$_{\it r}$ image should contain less or no scattered light from the disk. 

  The disk components are clearly visible in the Stokes {\it Q}$_{\it r}$ image, whereas the Stokes {\it U}$_{\it r}$ image does not show any circular structures and its signals are faint and noisy (Figure \ref{piqrur}c and d). Therefore, we concluded that the final PI$_{sub}$ image is robust.

\section{RESULTS}

\begin{figure} 
\begin{center}
\FigureFile(80mm,){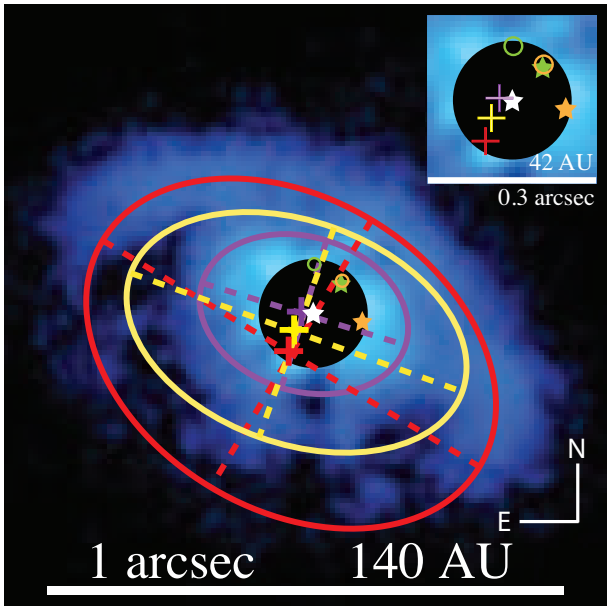}
\end{center}
\caption{Elliptical fitting results of the inner disk ({\it purple}), the gap ({\it yellow}), and the outer disk ({\it red}). The image has been smoothed by a gaussian with r=2 pixels to reduce the effects of speckles on the inferred structure of the disk. The central region is also shown in the right top panel. {\it White star} indicates the location of LkCa 15. {\it Green and orange stars} indicate where the planet candidates LkCa 15 b and c were detected in 2014, respectively (\cite{sal15}). {\it Empty green and orange circles} indicate the locations of two infrared sources seen in 2009-2010 (\cite{kra12}), which are assumed as LkCa 15 b and c, respectively.}\label{ell}
\end{figure}
  
    The final PI image of the LkCA 15 disk with a software mask (r$\sim$0.1$\arcsec$) is shown in the right panel of Figure \ref{piqrur}b and \ref{ell}, and two elliptical disk structures are clearly resolved. The brightness of the northwest side is significantly brighter than that of the southeast side, and this characteristic crescent of brightness is consistent with the optical imaging results of \citet{tha15}.

  The elliptical shape could be due to the system's inclination ({\it i}). Thus, we fitted elliptical isophotes on a resulting image in order to measure the inclinations and position angles (PAs) of each disk. The elliptical fitting results are shown in Figure \ref{ell} and Table \ref{tab}. We discovered new significant misalignments from major axis PAs of the two disks and gap (13$\pm$4$\degree$). The center of all three disk components appear on southeast side from the central star. The inclinations of the two disks are similar ($\sim$44$\degree$), but that of the gap shows larger angle ($\sim$52$\degree$). Note that \citet{tha14} reported eccentricities from the shape of the gap associated with LkCa 15, thus the inclination based on the ellipse fit only could be biased.

\begin{figure} 
\begin{center}
\FigureFile(80mm,){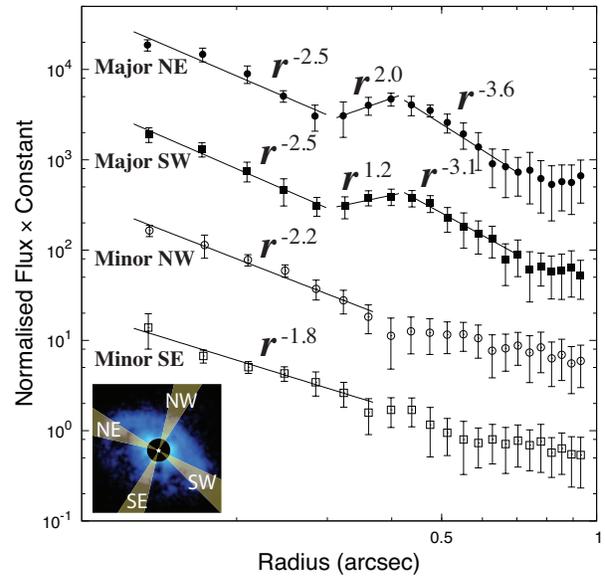}
\end{center}
\caption{Radial surface brightness profiles with 1$\sigma$ error bars at major and minor axes. The values were measured at each axis within the range of $\pm 10 \degree$ (Yellow regions in left bottom panel), and were binned with a width of {\it d}r=4 pixels. The typical error of of power index is $\sim$0.05.}\label{rad}
\end{figure}

Figure \ref{rad} shows the radial surface brightness profiles on the major and minor axes with a power-law fit at each slope. In the profiles of major axes (top two profiles in Figure \ref{rad}), the gap appears as a depletion in the middle of each profile. The slopes of the gap regions in profiles show a significant change between northeast and southwest axes (power indices $r$=2.0 and 1.2, respectively), and the slopes of the disk regions also show a change between inner and outer disks (r$=$-2.5 for inner disks, $r$=-3.1 and -3.6 for outer disks).

\begin{table}
  \tbl{Elliptical fitting results of two disks and gap$^a$.}{
  \begin{tabular}{cccc}
\hline
Parameter & Outer Disk & Gap & Inner Disk \\
\hline\hline
r$_{major}$ (AU) & 59.0$\pm$1.4 & 48.3$\pm$0.7 & 29.8$\pm$2.0 \\
PA$_{major}$ ($\degree$)$^b$ & 59$\pm$2 & 67$\pm$3 & 72$\pm$2 \\
$i$ ($\degree$)$^c$ & 44$\pm$1 & 51$\pm$2 & 44$\pm2$ \\
Center (mas)$^d$ & (-37$\pm$4,-83$\pm$6) & (-24$\pm$6,-42$\pm$6) & (-13$\pm$2,-8$\pm$2) \\  
\hline
\end{tabular}}\label{tab}
  \begin{tabnote}
$^a$ The peak, bottom, and half maximum positions (for the outer disk, gap, and inner disk, respectively) were obtained first from the radial profile at position angles every 10$\degree$. Then we conducted an elliptical fit by using the non-leaner least-squares Gauss-Newton algorithm with five free geometric parameters.\\
$^b$ Counterclockwise from north axis. \\
$^c$ Derived from the ellipticity. The inclination of a face-on disk is 0$\degree$, and that of an edge-on disk is 90$\degree$. \\
$^d$ ($\Delta$R.A., $\Delta$Dec.). The origin of the coordinate corresponds to the position of the central star.

  \end{tabnote}
\end{table}

\section{DISCUSSION}

\subsection{Disk Geometry: Which side is near to us?}

We revisited the question of which side of the disk faces us. The brightness asymmetry of the disk could be a clue, but two explanations can be provided for that. The first is backward illumination of the gap wall; backward illumination indicates that the bright side is the wall of the far side of the gap (\cite{qua11}). The second is forward scattering, which indicates that the bright side is the surface of the disk's near side (e.g., \cite{fuk06}). If backward illumination is the true explanation, the outer disk's inner edge would be optically thick and vertically high enough to conceal and reflect backward the light from the star. On the other hand, if forward scattering is the true explanation for this asymmetry, the inner edge of the outer disk would have a relatively low vertical height; therefore, more star light would arrive on the disk surface over the gap wall and more scattered light would come toward the observer. Therefore, both explanations are still under debate.

  To try to elucidate which side faces us, we utilized the star-disk offset along the minor axis. On the projection of inclined disk, the central star comes to near side of the disk's minor axis (\cite{don12}). In Figure \ref{ell}, the central star is roughly on the northwest minor axes of three disk components, therefore we can conclude that the northwest side of the disk of LkCa 15 could be the one facing us. This supports forward scattering as the explanation of the brightness asymmetry and is also consistent with 3D radiative transfer modeling \citep{tha14}.

\subsection{Surface brightness behavior}

We found the slope changes between northeast and southwest gaps, and between inner and outer disks. Furthermore, brightness slopes are not consistent with those of SPHERE/ZIMPOL results (power indices of inner disk, gap, and outer disk $\sim$ -2.4, 3.2 and -3.5 for northeast axis, -1.9, 2.1, and -4.1 for southwest axis, respectively, with typical error $\sim$ 0.1; \cite{tha15}, private communication).

A number of reasons have been suggested to explain the change in brightness slopes, such as different extinction levels, surface densities, flaring angles or dust properties. The geological properties of inner disk could change the extinction level between the star and outer region and affect the brightness of outer disk  (e.g., \cite{kri00}); an actual change in the surface density slope can be translated into a change in the surface brightness slope, and a change of the flaring angle cause a change in the scattering of the disk's surface (e.g., \cite{apa04}); a radial distribution of small dust particles and dust properties can affect the brightness slope (e.g., \cite{aki15}). Although the brightness behavior could provide some physical properties of the disks, a detailed analysis on the reason of the brightness behavior is out of the scope of this letter, and it will be discussed elsewhere.

\subsection{The origin of large gapped and warped disk}

In the PI$_{sub}$ image and radial profiles, LkCa 15 has a large gapped (width $\sim$27 AU) disk. Among some mechanisms (e.g., grain growth, photoevaporation, disk-planet interaction; see \cite{esp14}) that have been proposed to explain the clearing of the gaps in transitional disks, only gravitational interaction between disks and orbiting {\it multiple} planets can clear a {\it large} inner gap of $\gtrsim$ 15 AU or more (\cite{zhu11}) and preserve optically thick inner disk. Furthermore, \citet{jua13} suggested a 1 $M_{\rm Jup}$ planet would create a similar size of outer gap edge at NIR and (sub-)mm wavelengths; conversely planets more massive than 1 $M_{\rm Jup}$ make different radial grain-size distribution in the dusty disk, and observations at different wavelengths capture different parts of grain-size distribution. Since the sizes of the outer gap edge of LkCa 15 are $\sim$50 AU in sub-mm (\cite{pie06}) and 48 AU at NIR (this work), a 1 $M_{\rm Jup}$ planet might create a gap around LkCa 15. By combining the upper mass limits of LkCa 15 companions, as \cite{tha10} suggested based on their imaging result, we concluded that assuming multiple planets with a mass of $\lesssim$1$M_{\rm Jup}$ could account for LkCa 15's large gapped disk with an outer gap edge similar in size at both NIR and (sub-)mm wavelengths.  

We found a significant misalignment between two position angles of inner and outer disks (=13$\pm$4$\degree$) which indicates that the inner disk is possibly warped along the disk major axis. If inner disk was also warped along the minor axis, we would see misaligned inclination. However, the inclination of inner disk is consistent with that of outer disk ($\sim$44$^\circ$). Warped disks, such as $\beta$ Pictoris (e.g., \cite{mou97}), AB Aurigae \citep{has11} and HD 142527 (e.g., \cite{mar15}), have been reported on several stars surrounded by transitional disks and debris disks. These warped inner disks may be explained by the gravitational perturbation from planets (e.g., \cite{mou97}). $\beta$ Pictoris, whose planetary mass companion $\beta$ Pictoris b has a similar inclination to and possibly responsible for the inner warped disk (e.g., \cite{lag12}), is a possible evidence for this scenario. Additionally, \citet{ahm09} also suggested the possibility of multiple planets in $\beta$ Pictoris to explain warped disks.

\

   To summarise, since LkCa 15 may possess multiple planets with a mass of $\lesssim$1$M_{\rm Jup}$ in the large gap, the warped inner disk could be the result of potential planets around LkCa 15. 

\section{CONCLUSION}

We have presented a warped inner component beyond the large gap from the LkCa 15 disk system revealed by angular differential imaging in the {\it H}-band with HiCIAO installed on a Subaru Telescope. We derived 13$\pm$4$\degree$ as the PA offset between the outer disk and the warped inner disk. This unique gap plus the warped disk configuration of the LkCa 15 system combined with the previous observations at mm and optical wavelengths indicates the existence of a multiple planetary system possibly composed of $\lesssim$1$M_{\rm Jup}$ planets on the solar system scale. To directly observe and reveal the origin and evolution of possible multiple planetary systems, future ground-based observations with the Extreme Adaptive Optics (ExAO) system such as the Subaru Coronagraphic ExAO (SCExAO; \cite{jov15}) are required.

\

We are grateful to an anonymous referee for providing many useful comments leading to an improved version of this letter. We gratefully thank the assistance of the Subaru telescope. This work makes use of data provided by SMOKA. MT is supported by Grant-in-Aid for Scientific Research (No.15H02063).

\end{document}